\documentstyle[prl,aps,multicol]{revtex} 
\input{epsf}
\begin{document} 
\draft 
\title{Role of a parallel magnetic field 
 in two dimensional disordered clusters \\
 containing a few correlated electrons} 

\author{Franck Selva and Jean-Louis Pichard} 
\address{CEA, Service de Physique de l'Etat Condens\'e,
Centre d'Etudes de Saclay, F-91191 Gif-sur-Yvette, France} 
\date{\today}
\maketitle

\begin{abstract} 
 
 An ensemble of $2d$ disordered clusters with a few 
electrons is studied as a function of the Coulomb energy to 
kinetic energy ratio $r_s$. Between the Fermi system (small $r_s$) 
and the Wigner molecule (large $r_s$), an interaction 
induced delocalization of the ground state takes place which is 
suppressed when the spins are aligned by a parallel magnetic field. 
Our results confirm the existence of an intermediate regime where 
the Wigner antiferromagnetism defavors the Stoner ferromagnetism and 
where the enhancement of the Land\'e $g$ factor observed in dilute 
electron systems is reproduced. 

\end{abstract} 
\begin{multicols}{2}

%
%++++++++++++++++++ EXPERIMENT ON QUANTUM DOTS AND RELATED THEORIES +++++++++ 
%
 The magnetization of quantum dots is the subject of recent 
 studies \cite{berkovits,quantum-dot-theory} since their Ohmic resistances 
 are measured as a function of a gate voltage in the Coulomb blockade 
 regime. The Pauli rule (electrons populating the orbital states of 
 a dot in a sequence of spin up - spin down electrons) leads to a 
 bimodal distribution of the conductance peak spacings. The possibility 
 of a spontaneous magnetization $S$ of their ground state due to  
 electron-electron interactions has been proposed to explain the 
 absence of such a distribution in the experiments. Since the 
 Coulomb energy to Fermi energy ratios $r_s$ are not too large in those 
 dots, the interactions are mainly described assuming the Hartree-Fock 
 (HF) approximation. A break-down of the Pauli rule is obtained if, by 
 placing electrons at higher orbitals, the gain in exchange energy exceeds 
 the loss in kinetic energy. This is the Stoner mechanism \cite{stoner} 
 which eventually gives a ferromagnetic ground state. The mesoscopic 
 fluctuations of $S$ are usually studied \cite{quantum-dot-rmt} assuming 
 some random matrix toy model valid in the zero dimensional limit. We 
 numerically study in this letter another limit where the $2d$ character 
 of the dot is kept and where the interactions are exactly taken into 
 account. Therefore, our study is directly relevant for quantum dots 
 created from dilute $2d$ electron gases and may give insights into the 
 physics of the dilute $2d$ electron gas (strongly correlated limit).

%
%++++++++++++++++++ 2d-MIT +++++++++ 
%
 
The dilute limit is opening a new frontier for experiment and theory.
$2d$ electron gases can be created \cite{abrahams,batlogg} 
in field effect devices at low carrier densities $n_s$ and large ratios 
$r_s \propto {n_s}^{-1/2}$. 
The dependence of the $2d$ resistance $R$ as a function of the temperature 
$T$ and the applied magnetic field $B$ has been studied up to large 
values of $r_s$. Wigner crystallization, which remains to be  
confirmed in the experiments, is seen in computer simulations for values of 
$r_s$ where a metal-insulator transition (MIT) is observed 
\cite{abrahams,yoon} . One has an insulator when $r_s$ is large and a 
metal when $r_s$ is smaller. The MIT occurs \cite{yoon} at $r_s \approx 10$ 
in disordered devices, at larger $r_s$ in cleaner devices. Since Anderson 
localization takes place when $r_s \rightarrow 0$, the metallic behavior 
of $R(T)$ observed for intermediate values of $r_s$ suggests the 
possible existence of an intermediate metallic phase between two insulating 
limits of different nature: a Fermi system of Anderson localized states 
on one side and a pinned Wigner crystal on the other side. The existence 
of such a phase is supported by exact studies \cite{bwp1,bwp2} of small 
disordered clusters where three different regimes are observed for 
the persistent currents. Nevertheless it remains unclear that the observed 
intermediate regime is not a simple finite size effect, and that it 
does correspond to a new kind of metal. An analogy with the physics  
of $^{3}He$ has been recently discussed \cite{spivak}.  

%
%++++++++++++++++++ Polarized limit and BWP +++++++++ 
%

Renormalization group equations \cite{finkelshtein} obtained 
for weak disorder and weak interaction underline 
the role of the spins and suggest that a $2d$ metal might exist. 
The behavior of $R$ when a parallel magnetic field $B$ is applied 
(no orbital effects) confirms the role of the electronic spins. 
For the intermediate $r_s$ where $R(T)$ behaves as in a metal, a 
positive magnetoresistance \cite{abrahams} saturating at a large value 
for $B > B_{sat}$ is observed. Small-angle Shubnikhov-de-Haas 
measurements \cite{vitkalov1} confirm that $B_{sat}$ is the necessary 
field to polarize the electrons completely.  Following ref. \cite{shashkin1}, 
$B_{sat} \propto (n_s-n_c)$ where $n_c$ is the density where the MIT takes 
place at $B=0$. This behavior is observed for densities $n_s$ corresponding to 
$ 3 < r_s < 10$ and gives $g(n_s) \propto n_s(n_s-n_c)^{-1}$ for the 
Land\'e $g$ factor. 
A shift of $n_c$  (defined from the $I-V$ characteristics) is also 
observed \cite{shashkin2}:  $n_c$ increases as a function of $B$ before 
saturating when $B>B_{sat}$. This typically gives for the critical ratios 
$r_s^c$: $r_s^c(B=0)/ r_s^c ( B > B_{sat}) \approx 1.2$.

%
%++++++++++++++++++ summary of the main results +++++++++ 
%

 Our study of an ensemble of small disordered clusters 
confirms that the intermediate regime observed \cite{bwp1} 
using spinless fermions remains when the spin degrees of freedom 
are included. When the electrons are not polarized by a parallel field $B$, 
the ground state begins to be delocalized by the interaction 
before forming the localized Wigner molecule. When the spins are 
aligned by $B$, this delocalization effect is suppressed and the 
crystallization threshold is shifted by a small amount compatible 
with the experimental observations \cite{shashkin2}. The 
polarization field $B_{sat} \propto r_s^{-2}$ found in our simulations 
is consistent with the law $B_{sat} \propto n_s$ found in 
Ref. \cite{shashkin1} in a similar interval of intermediate values 
for $r_s$. Magnetic signatures of the intermediate regime are given. 
The local magnetic moments ($S=1$) appearing above the mesoscopic Stoner 
threshold $(r_s \approx 0.35$ for the studied clusters) begin to disappear 
above a first threshold $r_s^{FS}\approx 2.2$ where they have a maximum 
probability. Above $r_s^{FS}$, the Stoner mechanism 
and hence the HF approximation break down, $B_{sat} \propto r_s^{-2}$ and 
the field $B$ necessary to create a local moment ($S=1$) in a $S=0$ 
cluster becomes independent of $r_s$. Those behaviors disappear 
around $r_s^{WS} \approx 10$ where the electrons crystallize and form 
an antiferromagnetic ($S=0$) Wigner molecule.
  
%%%%%%%%%%%%%%%%%%%% Hamiltonian %%%%%%%%%%%%%%%%%%%%%%%%%%%%%%%%%%%%%%%%%%%

 The clusters are described by an Hamiltonian ${\cal H}$ of $N$ electrons 
free to move on a $L \times L$ square lattice with on-site disorder 
and periodic boundary conditions (BCs).
\begin{eqnarray}
{\cal H}&=& \sum_{i,\sigma} (-t \sum_{i'} c^{\dagger}_{i',\sigma}
c_{i,\sigma} + v_i n_{i,\sigma}) 
\nonumber \\
& &+\frac{U}{2} \sum_{i,i'\atop i\neq i'} 
\frac{n_{i,\sigma} n_{i',\sigma'}}
{|i-i'|}+ 2U 
\sum_{i} n_{i,\uparrow}n_{i,\downarrow}.
\end{eqnarray}
The operators $c_{i,\sigma}$ ($c^{\dagger}_{i,\sigma}$) destroy 
(create) an electron of spin $\sigma$ at the site $i$ and 
$n_{i,\sigma}=c^{\dagger}_{i,\sigma}c_{i,\sigma}$.
The hopping term $-t$ couples nearest-neighbor sites
and the random potentials $v_i$ are uniformly distributed inside 
$[-W/2,W/2]$ with $W=5$ (diffusive regime for non interacting 
particles). Two electrons on the same site costs a $2U$ Hubbard energy 
and a $U/|i-i'|$ Coulomb energy  if they are on two sites $i$ and $i'$ 
separated by $|i-i'|$ (shortest length separating $i$ and $i'$ on a 
square lattice with periodic BCs). The use of the dimensionless ratio 
$r_s=U/(2t \sqrt{\pi n_s})$ where $n_s=N/L^2$ allows us to compare our 
results obtained as a function of $U$ and the experimental results 
obtained as a function of $n_s$. 

%%%%%%%%%%%%%%%%%%% Determination of S %%%%%%%%%%%%%%%%%%%%%%%%%%%%%%%%%%%%%%

We denote ${\cal S}$ and ${\cal S}_z$ the total spin and its component 
along an arbitrary direction $z$. Since $[{\cal S}^2,{\cal H}]=
[{\cal S}_z,{\cal H}]=0$, ${\cal H}$ can be written in a block-diagonal 
form, with $N+1$ blocks where $S_z=-N/2,\ldots,N/2$ respectively. The blocks 
with $S_z$ and $-S_z$ are identical in the absence of a magnetic field $B$. 
When $B=0$, there is no preferential direction and the groundstate energy 
$E_0$ does not depend on $S_z$. For a groundstate of total spin $S$, 
${\cal H}$ has $2S+1$ blocks with the same lowest eigenenergy $E_0 (S^2)$ 
since $ E_0(S^2)=E_0(S^2,S_z); \; \; \scriptstyle{S_z=-S,-S+1,...,S-1,S}$. 
Therefore, the number $N_b$ of blocks of different $S_z$ and of 
same lowest energy gives the total spin $S=(N_b-1)/2$ of the groundstate. 
We consider $N=4$ and $L=6$, as in ref. \cite{bwp1}. 
$S_z = 2,1,0,-1,-2$ give  two characteristic 
energy differences $\Delta_1 = E_0(S_z=1)-E_0(S_z=0)$ and 
$\Delta_2=E_0(S_z=2)-E_0(S_z=0)$. The $E_0(S_z)$ energies are exactly 
obtained using Lanczos algorithm. The size of the matrices to diagonalize 
are $N(S_z)=396900, 257040$ and $58905$ for $S_z=0,1,2$ respectively.

 A parallel magnetic field $B$ defines the $z$-direction and removes 
the $S_z$ degeneracy by 
the Zeeman energy $-g\mu B S_z$. The ground state 
energy and its magnetization are then given by the minimum of 
$E_0(S^2,S_z,B=0)-g\mu B S_z $. For a $S=0$ groundstate without field, 
the value $B^*$ for which $E_0(S_z)-g\mu B^* S_z \textstyle{=E_0(S_z=0)}$ 
defines the field  necessary to polarize the system 
to $S\geq S_z$. The total $\Delta_2$ and partial $\Delta_1$ polarization 
energies are the Zeeman energies necessary to yield $S=2$ and $S=1$ 
respectively for a cluster with $S=0$.  

%%%%%%%%%%%%%%%%%%%%%%% U=0 and Large U limit %%%%%%%%%%%%%%%%%%%%%%%%%%%%%%

 When $r_s=0$, the two one body states of lowest energy are doubly occupied 
and $S=0$ ($S=1/2$ if $N$ is odd). To polarize the $S=0$ ground state to 
$S=1$ corresponds to the transition of one electron at the Fermi energy 
and costs an energy equal to the one body level spacing. $p(\Delta_1)$ 
is then given by the spacing distribution $p(\delta)$ between consecutive 
one body levels, the Wigner surmise $(\pi \delta/2) \exp -(\pi \delta^2/4)$ in 
the diffusive regime. $\Delta_2$ corresponds to the sum of a few one body 
excitations. 

 When $r_s$ is large, the $4$ electrons occupy the four sites 
$c(j)$ $j=1,\ldots,4$ of the square configuration of side $a=3$  
with the lowest substrate energy $\sum_{j=1}^4 v_{c(j)}$. If  
$|0>$ denotes the vacuum state, the ground state 
in this limit becomes $|\Psi_c>=\prod_{j=1}^4 c^{\dagger}_{c(j),\sigma_j} |0>$ 
with a spin independent energy $E_c$. This square can support $2^N=16$ 
spin configurations. We summarize the main results of a perturbative 
expansion of $E_c$ in powers of $t/U$. The spin degeneracy of 
$E_c$ is removed by terms of order $t(t/U)^{2a-1}$, which is the 
smallest order where the $16$ spin configurations can be coupled via 
intermediate configurations allowing a double occupancy of the same 
site. Therefore, $2a-1$ is the order where the perturbation begins 
to depend on $S_z$ and $\Delta_1$ as $\Delta_2 \propto t(t/U)^{2a-1} 
\rightarrow 0$ when $t/U \rightarrow 0$ (we have numerically checked 
this decay when $r_s > 100$). Moreover, the correction  to $E_c$ 
depending on $S_z$ and $\propto t(t/U)^{2a-1}$  is given by an effective 
antiferromagnetic Heisenberg Hamiltonian. The $S=0$ ground state for large 
$r_s$ correspond to $4$ electrons forming an antiferromagnetic square Wigner 
molecule. This does not reproduce the ferromagnetic ground state suggested 
in Ref. \cite{vitkalov2}. However $\Delta_2$ is very small 
when $r_s$ is large, and the antiferromagnetic behavior can be an artefact 
due to the square lattice. Without impurities, a quasi-classical WKB 
expansion \cite{roger} shows that 3 particle exchanges dominate in the 
continuous limit, leading to ferromagnetism. Recent Monte-Carlo calculations 
\cite{bernu} suggest that the crystal becomes a frustrated antiferromagnet 
closer to the melting point.

%%%%%%%%%%%%%%%%%%% Distribution of polarization %%%%%%%%%%%%%%%%%%%%%%%%%%%%   
The perturbative corrections $\propto t(t/U)^{2a-1}$ depend 
on the random variables $v_i$ via $\prod_{J=1}^{2a-1} (E_c-E_J)^{-1}$ 
where the $E_J$ are the classical energies of the intermediate configurations. 
$E_J$ is the sum of an electrostatic energy and of a substrate energy 
$E_s(J)=\sum_{k=1}^4 v_{J(k)}$. Due to the high order $2a-1$ of the 
correction, a normal distribution for $E_J$ leads to a log-normal 
distribution for $\prod_{J=1}^{2a-1} (E_c-E_J)^{-1}$. Therefore 
$p(\Delta_1)$ and $p(\Delta_2)$ should be log-normal when $r_s$ is large.  

\vspace{-0.5cm}
\begin{figure}
\centerline{
\epsfxsize=8cm
\epsfysize=8cm
\epsffile{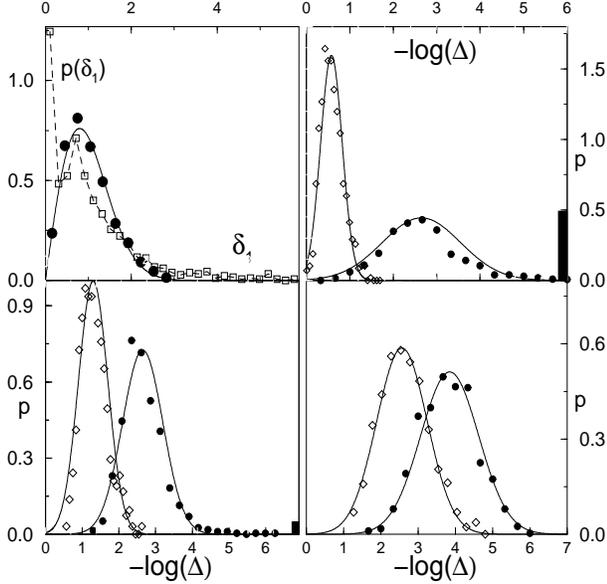}
}
\caption
{
Distributions of $\Delta_1$ and $\Delta_2$ at different values of $r_s$. 
Upper left: $p(\delta_1)$ at $r_s= 0$ (circle) and 
$2.5$ (square) where $\delta_1=\Delta_1/<\Delta_1>$. 
The continuous line is the Wigner surmise.  
$p(-\log \Delta_1)$ (circle) and $p(-\log \Delta_2)$ (diamond) 
at $r_s=2.5$ (upper right) $ 5.8 $ (lower left) and $16.8$ (lower right) 
respectively. The thick bars (put at right edge of the figures) 
give the peaks $\delta(\Delta_1)$ of the bimodal $p(\Delta_1)$. 
The continuous lines are normal fits.
}  
\label{fig1bis}
\end{figure}

 $p(\Delta_1)$ is given in Fig. \ref{fig1bis} for different $r_s$.  
The expected Wigner surmise takes place for $r_s=0$. 
A small interaction quickly drives $p(\Delta_1)$ towards a bimodal 
distribution, with a delta peak at $\Delta_1=0$ 
and a main peak centered around a non zero value of 
$\Delta_1$. The delta peak gives the probability to have spontaneously 
magnetized clusters with $S=1$. The main peak gives the field $B$ 
necessary to create $S=1$ in a cluster with $S=0$.  
The logarithmic scale used in Fig. \ref{fig1bis} underlines  
the bimodal character of the distribution and confirms that the main peak 
becomes log-normal when $r_s$ is large. The distribution of $\Delta_2$ is not 
bimodal: a fully polarized cluster has never been seen when $B=0$. 
$\Delta_2$ becomes also log-normally distributed when $r_s$ is large.  

\begin{figure}
%\vspace{-1cm}
\centerline
{
\epsfxsize=8cm
\epsfysize=8cm
\epsffile{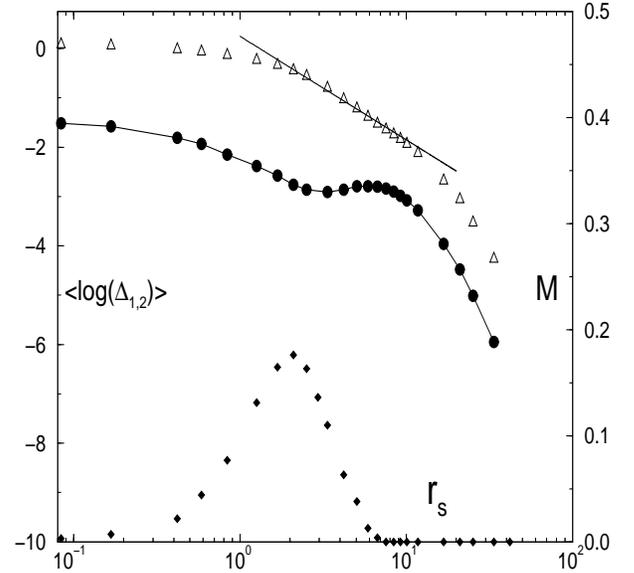}
}
\vspace{0.5cm}
\caption
{
As a function of $r_s$, fraction $M$ (diamond) of clusters with $S=1$ 
at $B=0$,  partial $<\log \Delta_1>$ (circle) and total 
$<\log \Delta_2>$ (triangle) energies required 
to polarize $S=0$ clusters to $S=1$ and $S=2$. 
The straight line corresponds to $0.25 - 2 \log r_s$. 
}
\label{fig2}
\end{figure}

 In Fig. \ref{fig2}, the fraction $M$ of clusters with $S=1$ at $B=0$ 
is given as a function of $r_s$. One can see the mesoscopic Stoner 
instability \cite{stoner} taking place at $r_s \approx 0.35$. The Stoner 
mechanism should eventually give fully polarized electrons. This is not 
the case, the increase of $M$ breaks down when $r_s= r_s^{FS} \approx 2.2$, 
a value where the Stoner mechanism and hence the HF approximation break down. 
In the same clusters, the HF approximation fails\cite{benenti} to describe 
the persistent currents of $4$ spinless fermions when $r_s> r_s^{FP} 
\approx 5$. $r_s^{FP}$ takes a smaller value $r_s^{FS}$ when the spin 
degrees of freedom are included. Above $r_s^{FS}$, $M$ regularly decreases 
to reach a zero value for $r_s^{WS} \approx 9$ where an antiferromagnetic 
square molecule is formed. In the intermediate regime, there 
is a competition between the Stoner ferromagnetism and the Wigner 
antiferromagnetism. Since the $S=0$ clusters are characterized by log-normal 
distributions, the ensemble averages $<\log \Delta_1>$ and 
$<\log \Delta_2>$ (without taking into account the $S=1$ spontaneously 
magnetized clusters) define the typical fields $B$ necessary to 
yield $S=1$ or $S=2$ in a $S=0$ cluster. Fig. \ref{fig2} provides 
two magnetic signatures confirming the existence of a novel intermediate 
regime between the Fermi glass $(r_s < r_s^{FS})$ and the Wigner glass 
$(r_s>r_s^{WS})$: $<\log \Delta_1>$ becomes roughly independent of $r_s$, 
while $\Delta_2 \propto r_s^{-2}$. Very remarkably, this $r_s^{-2}$ dependence 
is consistent with the $n_s$ dependence of $B_{sat}$ seen in the 
experiments \cite{shashkin1}. 

\vspace{-0.5cm}
 
%%%%%%%%%%%%%%%%%%%%%  Magnetotransport %%%%%%%%%%%%%%%%%%%%%%%%%%%%%%%%%%%%

We now study three quantities calculated from the three 
eigenstates $|\Psi_0(S_z)>$ of lowest energy with 
$S_z=2,1,0$ respectively: (i) the number $\xi (S_z) = 
N^2(\sum_i\rho_i^2(S_z))^{-1}$ of occupied lattice sites 
($ \rho_i(S_z)=<\Psi_0(S_z)|\sum_\sigma n_{i,\sigma}|\Psi_0(S_z)>$);  
(ii) the crystallization parameter $ \gamma(S_z)=max_r C(r,S_z)-min_r 
C(r,S_z))$ where $C(r,S_z)=N^{-1}\sum_i \rho_i(S_z)\rho_{i+r}(S_z)$. 
$\gamma(S_z)=0$ ($1$) if the state is a liquid (a crystal); 
(iii) the longitudinal $I_l(S_z)$ and transverse $I_t(S_z)$ 
components of the total persistent current driven by an Aharonov-Bohm 
flux $\phi=\pi/2$ enclosing the $2d$ torus along the longitudinal 
direction ($\phi=\pi$ corresponds to anti-periodic longitudinal 
BCs and periodic transverse BCs, see Ref. \cite{bwp2}). 

\begin{figure}
\centerline{
\epsfxsize=8cm
\epsfysize=8cm
\epsffile{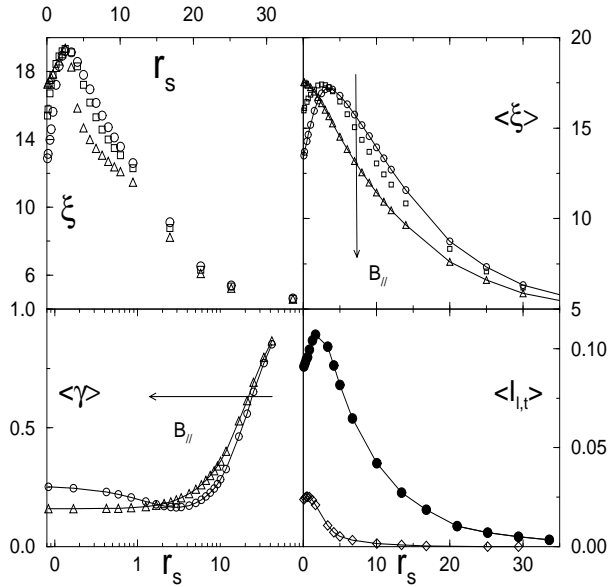}
}
\caption
{
As a function of $r_s$, numbers  $\xi (S_z)$ of occupied sites for a 
typical sample (upper left); ensemble averages $<\xi (S_z)>$ (upper right) 
and crystallization parameter $<\gamma (S_z)>$ (lower left) 
for the lower energy states with $S_z=0$  (circle), 
$S_z=1$  (square) and $S_z=2$  (triangle);  Average longitudinal 
$<I_l(S_z=0)>$ (circle) and transverse (diamond) $<I_t(S_z=0)>$ 
components of the total persistent current (lower right).
}
\label{fig3}
\end{figure}

$\xi(S_z)$ (shown for a single cluster in Fig. \ref{fig3})
depends on $S_z$ for small $r_s$ and becomes independent of $S_z$ 
for large $r_s$. At $r_s=0$, $\Psi_0(S_z=2,1,0)$ occupy respectively 
$4,3,2$ one body states while the Wigner molecule occupies $4$ sites 
only at large $r_s$. The ensemble average $<\xi(S_z=2)>$ is maximum 
when $r_s=0$ and decays as $r_s$ increases, suggesting the absence of 
delocalization for the polarized system. The non polarized system 
behaves differently, since $<\xi(S_z=1,0)>$ first increase to reach a 
maximum $\approx <\xi(S_z=2,r_s=0)>$ before decreasing. 

In the lower left 
part of Fig.\ref{fig3} one can see that charge crystallization is easier 
when the clusters are polarized than otherwise. The shift of the 
critical threshold $r_s^W$ is consistent with the shift of the critical 
density reported in Ref. \cite{shashkin2}. The arrows indicated in 
Fig. \ref{fig3} underline two consequences of a parallel field $B$: 
smaller number of occupied sites and smaller crystallization threshold. 
This is qualitatively consistent with the large magnetoresistance observed 
in the metallic phase for intermediate $r_s$. Fig. \ref{fig3} 
(lower right) gives the total longitudinal $<I_l(r_s)>$ and transverse 
$<I_t(r_s)>$ currents for $S_z=0$. When the spin degrees of 
freedom are included, the conclusions previously obtained for spinless 
fermions \cite{bwp1,bwp2} remain valid: there are a Fermi regime where 
longitudinal and transverse currents coexist, an intermediate regime 
where the transverse current is suppressed while the longitudinal current 
persists, and a Wigner regime with vanishing persistent currents \cite{sw}. 
   
 In summary, our study suggests that Coulomb repulsion by itself provides 
a qualitative mechanism able to give together a delocalization of the 
electrons (``metallic behavior'') and a large positive magneto-resistance 
for intermediate value of $r_s$. The shift of the critical threshold as a 
function of $B$ and the dependence of $B_{sat}$ as a function of 
$r_s$ are reproduced. Two important questions remain opened. 
Do the finite size behaviors which we have observed persist in the 
thermodynamic limit? What are the physical explanations of the interplay 
between the Stoner ferromagnetism and the Wigner antiferromagnetism 
seen for intermediate values of $r_s$. An intermediate regime where a floppy 
antiferromagnetic Wigner molecule with vacancies coexists \cite{katomeris} 
with a ferromagnetic liquid of intersticial particles is suggested as a 
possible explanation.

%%%%%%%%%%%%%%%%%%%%%%%%%%%%%%%%%BIBLIO%%%%%%%%%%%%%%%%%%%%%%%%%%%%%%%

\end{multicols} 
\end{document}